\documentclass[aps,prl,a4paper,twocolumn,showpacs,superscriptaddress]{revtex4}
\usepackage{graphics}
\usepackage{epsfig}
\usepackage{color}
\usepackage[english]{babel}
\usepackage{natbib}

\usepackage{fancybox}
\begin{document}
\title{Do Bitcoins make the world go round? \\ On the dynamics of competing crypto-currencies}
\author{S. Bornholdt}
\affiliation{Institute for Theoretical Physics, University of Bremen, 28359 Bremen, Germany}
\author{K. Sneppen}
\affiliation{Niels Bohr Institute, Blegdamsvej 17, DK-2100 Copenhagen, Denmark}

\begin{abstract}
Bitcoins have emerged as a possible competitor to usual currencies, but other crypto-currencies
have likewise appeared as competitors to the Bitcoin currency. 
The expanding market of crypto-currencies now involves capital equivalent to $10^{10}$ US Dollars, 
providing academia with an unusual opportunity to study the emergence of value. 
Here we show that the Bitcoin currency in itself is not special, but may rather
be understood as the contemporary dominating crypto-currency that may well be replaced by other
currencies. We suggest that perception of value in a social system is generated by a 
voter-like dynamics, where fashions form and disperse even in the case where information 
is only exchanged on a pairwise basis between agents.  
\end{abstract}
\pacs{89.65.-s, 05.50.+q, 05.65.+b, 64.60.De}

\maketitle

\section{Introduction}

The recent surge in interest and value of Bitcoins is fuelled by lack of 
confidence in the usual banking system and its lack of transparency.
Currencies issued by central banks are not conserved and printing money 
has been a frequent response to various fiscal problems, abundantly spanning, 
both, nations and centuries. In recent years, the financial crisis led to 
bank bailouts and financial rescuing of whole nations, all at the expense 
of the big central currencies. Even private assets in banks were considered 
to be devaluated for the purpose of rescuing a national financial system. 
In this context, the prospects of a peer-to-peer currency without the 
need for a central bank meets the desire of many people. The 
crypto-currency Bitcoin, within only 5 years, has reached a market 
capitalization equivalent to ten billion US Dollars and therefore proves 
to be a popular new medium. 

\begin{figure}[h]
\centerline{\epsfig{file=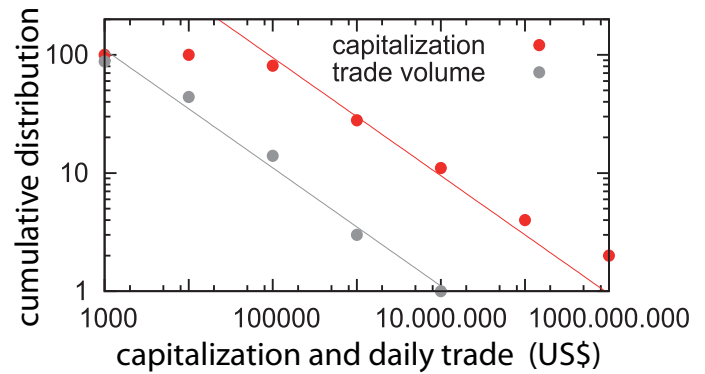,angle=0,width=9cm}}
\caption{\small\sl Estimated value of the 100 highest valued crypto-currencies,
using data from crypto-currency market capitalization from March 18, 2014
(coinmarketcap.com).
Also the figure shows the distribution of trade volumes at that day.
Both plots are cumulative, and exhibit a scaling over 2.5 orders of magnitude
with an exponent of about -0.5 (straight line fits). 
Although the Bitcoin currency is by far the best 
known crypto-currency, it does not distinguish itself from the overall distribution 
of important crypto-currencies as, for example, Litecoins or Dogecoins.}
\label{figure1}
\end{figure}

A major property of Bitcoins is its built-in limitation to a finite number of 
currency units, called coins. 21,000,000 coins in total can be generated, 
not more. This is in contrast to common currencies that can be printed 
in secret by central banks or can be devaluated by excessive issuing of loans. 
Cryptrographic methods ensure full transparency of the absolute conservation 
of Bitcoins and are therefore a considerable source of trust into this new medium. 

The large volatility of the Bitcoin currency, as well as its 
low number of daily transactions and low trading volume, support arguments 
that Bitcoins do not yet share the characteristics of a mature currency. 
And, furthermore, a major caveat has become clear in recent months: 
While the number of coins in the new currency is conserved, the overall 
number of crypto-currencies is not. As the underlying software is open source, 
cloning a crypto-currency is an easy matter, as is the release of a 
modified version of a crypto-currency. Today, hundreds of crypto-currencies 
can be found on the World Wide Web, and an unknown large number may be 
waiting for acceptance online. 

As a result, the constant volume of Bitcoins faces an unlimited number 
of alternative crypto-currencies and, therefore, an unlimited number of 
alternative coins. It is an interesting question whether this neutralizes 
the advantage of the finite inventory in Bitcoins. Clearly, an investor 
may move his assets from Bitcoins to a competing currency, thereby freely 
moving in a space with an unlimited number of coins. A quick look at 
a current crypto-currency exchange shows that so far Bitcoin capitalization 
dominates. 

Fig.\ \ref{figure1} shows that the relative strength of the 100 most valued
crypto-currencies in fact does not distinguish Bitcoin as special.
Rather, both the total market capitalization and the number of trades
vary as a power law, with the number of currencies exceeding $M$
decrease as $1/M^{0.5 \pm 0.1}$. Thus the major advantage of the Bitcoin
is its historical position, but it could in principle as well be replaced 
with any of its competitors.

Here we propose to view the value of any crypto-currency from a popularity 
standpoint, where coins gain foothold in a market because people communicate 
about the currencies and thereby act according to the currencies' popularity.
The simplest process that takes such dynamics into account is the Moran process
from evolutionary biology \cite{Moran58}, often rephrased as the Voter Model in 
Sociophysics \cite{Clifford73,Liggett75,Liggett97}.

\section{Model and Results}

\begin{figure}[h]
\centerline{\epsfig{file=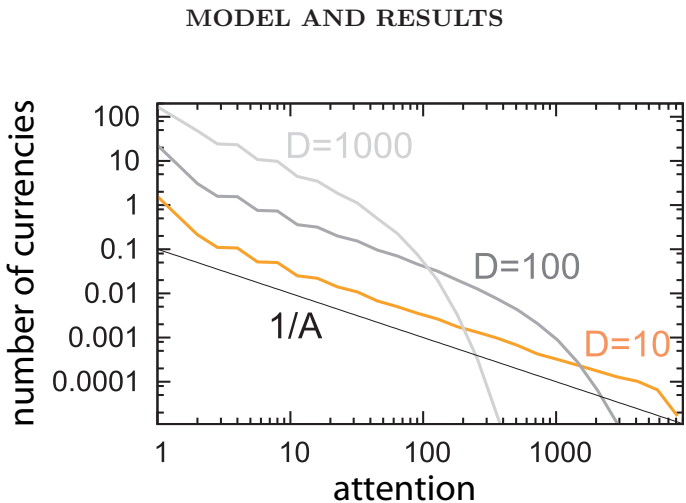,angle=0,width=9cm}}
\caption{\small\sl Distribution of popularity in steady state of
an iterated voter- or Moran process, where mutually exclusive states 
compete for N=10000 different sites. 
At each time-step one chooses two sites at random and copies the state of one 
onto the other site. In addition, each time a state is eliminated from the system, 
another state is introduced by inserting it at a random site, 
thereby replacing the previous state of that site. 
The three curves refer to a constantly maintained diversity of states of 
$D=10$ (orange), $D=100$ (dark grey),  and $D=1000$ (light grey), 
respectively. The straight black line has a slope of $-1$. 
}
\label{figure2}
\end{figure}

In terms of multiple currencies, a fashion process may be modeled as a number of 
sites, each representing memory slots in one person's mind. Naturally,
there may be more than one memory slots per person, representing the possibility
that this person knows about several crypto-currencies, and also that this person
perhaps allocates more memory to Bitcoins than for example to Litecoins.
The update of this ``fashion model'' mimicks communication between two people,
allowing one person to swap the state of one memory slot with the state of 
another memory slot from the other person.
When a particular coin is not present in any memory slot of any person,
it is considered to be eliminated. To generate a steady state process we here 
allow invention of a new coin by introducing it into one memory slot.
The process is simulated for a $N=10000$ system 
with a number of crypto-currencies $D=10$, giving a time averaged 
distribution of popularity $\propto 1/x$, see Fig.\ \ref{figure2}.

The above copying process is a classical model from sociology, implicitly 
suggested by Spencer in 1855 by his famous statement of proportionality 
between importance and how often people hear about a given subject
\cite{Spencer1855}.
From comparing the probability density Fig.\ \ref{figure2} with 
the cumulative plot in Fig.\ \ref{figure1} we see qualitative agreement,
although the capitalization of real crypto-currencies is decreasing faster with 
size than the model expectation. 

Based on the Moran process above, we now want to model a more complete market 
where agents also trade currencies and where crypto-currencies are mined
at a constant rate. The latter mining aspect will be implemented 
to mimic the long period of Bitcoins where a constant amount of about 
50.000 coins has been mined per week.
Our agent based model will remain simplified in the sense that it only 
considers market fluctuations related to peer to peer exchange, 
and further by considering only steady state properties. 

\begin{figure}[h]
\centerline{\epsfig{file=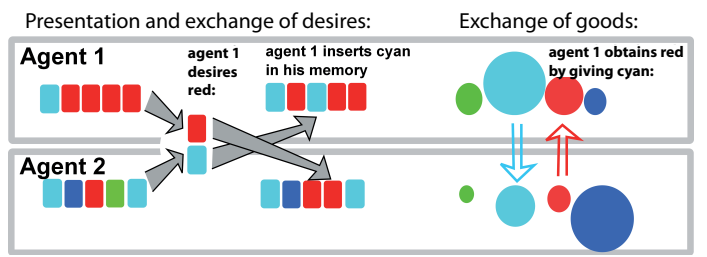,angle=0,width=9cm}}
\caption{\small\sl Two basic elements in the exchange between
two randomly selected agents. First, on the left the two agents
present one of their desires each, based on their allocated memories.
The presented desire makes the opposing agents replace
a fraction of their memories accordingly 
\cite{Yasutomi95,Donangelo02}, thereby
mimicking the Voter/Moran-like process simulated in Fig.\ 2.
Second, on the right, agents may exchange goods, based
on their respective perception of the value of these goods.
In our suggested market dynamics we implement the value of a good as 
being simply equal to the fraction of memory that an agent 
has allocated to this good.  
}
\label{figure3}
\end{figure}

The model describes a market of $N$ agents that invest, mine, and trade in $D$ 
different crypto-currencies. The trades will occur between agents 
on a one-on-one basis, always involving information exchange and 
occasionally also exchange of coins. Each agent $i$ 
has two types of internal variables. 
First a repository for assets, 
implemented as a vector $a_{i}(k)$ containing the respective 
assets of currencies $k=1,...,D$. 
A second vector $m_{i}(j)$, $j=1,...,mem$ contains 
the preferences of the agent in future acquisitions, with index number $j$ 
containing a memory $m_i(j)$ referring to one of the $D$ currencies.
Below we also use the derived quantity $M_{i}(k)$ that counts
the number of times currencies $k=1,2...D$
appear in the memory list of agent $i$. 
The model is defined in discrete trade steps:
\begin{itemize}
\item
{\bf Communication:} Select two different agents $i_1$ and $i_2$ 
and one memory slot from each ones' memory, subsequently referred to 
as currency $c1$ and currency $c2$. 
Let agents communicate by replacing a random one of their own interest 
slots with the selected interest slot from the other agent. 
If one currency is not any more present in the memory of any agent, 
it is eliminated from the system and a new currency is introduced in
its place with one coin unit and one memory slot of a randomly selected agent.
\item
{\bf Trades:} If agent $i_1$ has coin $c2$ in his inventory list 
and agent $i_2$ has coin $c1$, then the two agents may perform a trade, 
provided that both believe to gain.
Agent $i_1$ evaluates the value of one $c1$ unit to be $M_{1}(c1)+1$, thereby also
setting a lower threshold for value of the currency.  
Similarly, agent $i_1$ evaluates the value of $c2$ as $M_{1}(2)+1$,
whereas agent $i_2$ estimates the value of $c1$ and $c2$ 
respectively to $M_{2}(1)$ and $M_{2}(2)$. 
A transfer of $y$ numbers of $c_2$ coins from $i_1 \rightarrow i_2$
and $x$ number of $c_1$ coins $i_2 \rightarrow i_1$ can take place if
$M_{1}(c1)/M_{1}(c2) > M_{2}(c1)/M_{2}(c2)$ with exchange rate
$p=y/x=\sqrt{M_{1}(c1)\cdot  M_{2}(c1)/(M_{1}(c2) \cdot M_{2}(c2))}$. 
At the exchange, the maximal possible exchange is taking place: 
If $a_{1}(c2)>p\cdot a_{2}(c1)$ then 
$x=a_{2}(c1)$ and $y=p \cdot a_{2}(c1)$. 
If $a_{1}(c2) < p \cdot a_{2}(c1)$ then $x=a_{1}(c2)/p$ and $y=a_{1}(c2)$.
\item
{\bf Mining:} With a small probability all currencies are mined,
and each agent $i$ increases its amount of currency $c$ at a rate 
proportional to the fraction that this currency fills the memory of $i$,
divided by the total memory of all agents allocated to this currency.
Thereby, each currency is mined at a constant rate, whereas individual agents
will find it much harder to mine popular currencies.  
\end{itemize}
Notice that trading deals with any fractions of coins,
and accordingly the overall behavior, is independent of absolute
numbers of coins in the game, including the mining rates in step 3.  
Also note that there is no feedback from the trade step to the 
updating dynamics of the memories, leaving the underlying fluctuation 
in popularity to be very close to the multi-species Moran process simulated 
in Fig.\ 2, implying that the overall dynamics of global memory of a 
particular currency will be close to a Moran process. 

\begin{figure}[h]
\centerline{\epsfig{file=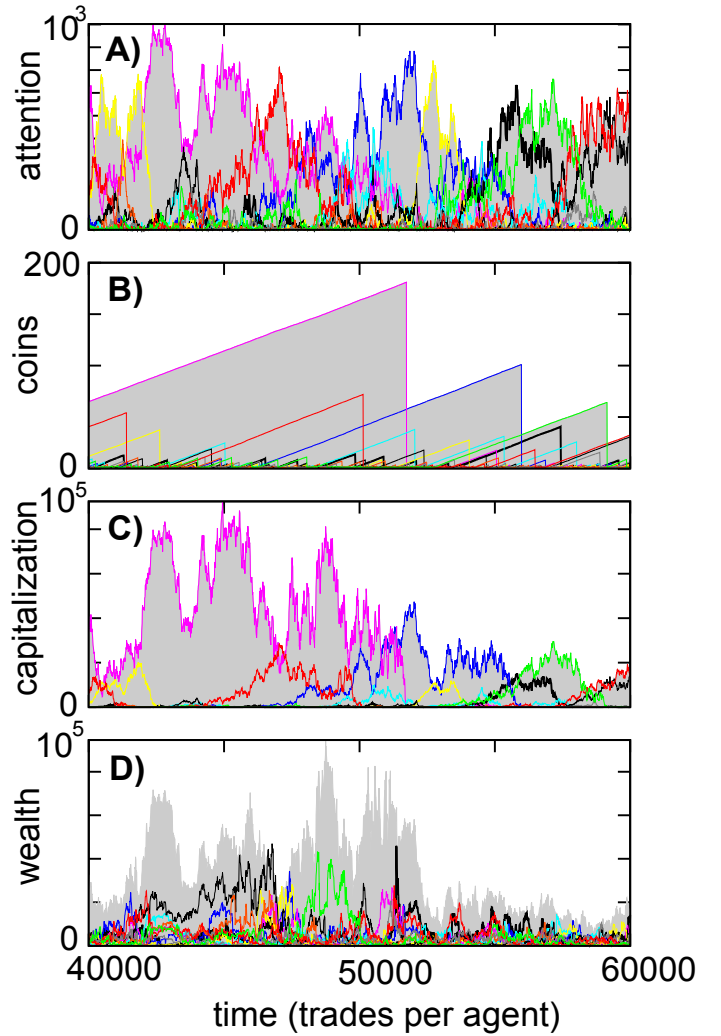,angle=0,width=9cm}}
\caption{\small\sl Dynamics of a model simulation with $N=100$,
a constantly maintained number of $D=10$ 
different crypto-currencies, and $mem=10$ memory slots per agent. 
a) Representation of the different currencies in the memories of the agents, 
measuring their popularity with time. 
b) The number of coins of various currencies,
as they get mined steadily during the lifetime of each crypto-currency.
A crypto currency is removed when it no longer appears in the memory
of any agent, and a new crypto-currency is then introduced by assigning
one coin and one memory unit into one random agent in the system.
c) Capitalization of each currency, defined as its total attention
multiplied by the total number of coins. 
d) Wealth of agents, using the trading scheme in the text where
goods are exchanged by using a local value of currency $i$ equal to 
$M_i+1$ where $M_i$ is the memory of the agent allocated to coin $i$. 
As in the previous plot, the grey area marks the most wealthy agent in the system.
}
\label{figure4}
\end{figure}

\begin{figure}[h]
\centerline{\epsfig{file=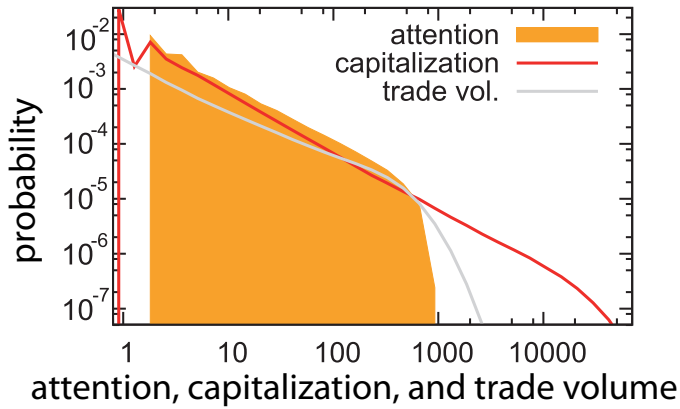,angle=0,width=9cm}}
\caption{\small\sl 
Distribution of attention and capitalization when sampling more than $10^5$
updates per agent in the system, and collecting data after a transient of 20000
time-steps. Parameters are the same as in the previous figure. 
Notice in particular that the scaling of capitalization is
close to that of the popularity (memory) of the currencies. 
}
\label{figure4}
\end{figure}

Fig.\ \ref{figure4} illustrates the typical behavior of the model,
with emphasis on the basic processes: attention, mining, capitalization and trading.
All time courses are shown in units of exchange between $N=100$ pairs of agents,
each with $mem=10$ memory slots, trading $D=10$ different crypto-currencies.  
In 
A) we see that dominance is not secured. 
Due to the ongoing introduction of new currencies
the memory allocated to also dominating currencies 
will be under constant challenge,
and ultimately the dominance is predicted to shift. 
At the same time, some currencies
get forgotten, to allow for new currencies with mining allowing their coin count to
increase linearly, as seen in panel B).
Panel C) shows the total capitalization of currencies,
a quantity that for each currency is equal to its total 
attention multiplied with its total coin count.
Panel D) finally shifts the focus to the agents' trading and mining the currencies,
a plot that illustrates that fluctuating currencies indeed open for a market
where people may get rich, but as well may get poor. The overall steady state
wealth distribution is found to be close to log-normal (not shown), 
as the individual agents' fortune is the result of a partly multiplicative 
process of success and failures. 
 
Fig.\ \ref{figure4} shows that the distributions of both attention (memory) 
and total capitalization remain close to the overall expectation of an 
iterated Moran process, i.e.\ $\propto 1/x$. Also the figure shows that trading 
volumes indeed follow a similar scaling, with large trades being associated 
to an exchange of popular currencies. 
Overall, the scale free distributions of real crypto-currencies are recapitulated,
with the caveat that our model shows a substantially broader distribution than the 
observed distributions. 

\section{Discussion}

This paper proposes to study the emergent crypto-currencies as a model system
of emerging and competing values. Emergence of value and money is an old 
and classical problem in economic literature, starting with Mengen \cite{Menger94} 
and later modelled through an interplay between need and fashion by 
\cite{Yasutomi95,Donangelo02}. In fact the above trading model shares
the communication step with \cite{Donangelo02}, but does not couple
the distribution of goods among agents to their value assessment,
a coupling that would expect to be more strategic than a need for diversification.
Crypto-currencies provide us with a fresh model system, presenting
a real world phenomena of value that has emerged without any 
need or fundamental value at all. The apparent rise of Bitcoins 
to the status of a currency which already now can be used to buy real 
products on the World Wide Web, thus indeed emphasizes that money is a social 
concept that can self-organize from simple contacts between people. 

Our model is at its core simplistic, re-iterating the basic fact that
all crypto-currencies are inherently interchangeable and well may 
be reshuffled by future contingencies. Our model fails to give 
the same exponent as observed for the power law scaling of real 
crypto-currencies, and instead predicts systematically broader distributions.  
Said differently, the contemporary dominance of Bitcoin is in fact less 
than one would typically expect of a voter- or Moran-like underlying 
social dynamics. 

Possible limitations of our model may be in particular
the possibility to obtain more dramatically fluctuating fashion dynamics,
caused by an interplay between global information spreading, marketing,
or including propagation of potentially catastrophic news. Thus other
types of preferential growth of attention may be considered, for example
the rich-gets-richer dynamics by Simon \cite{Simon55} with an overall
prediction of a wealth distribution with scaling exponent steeper 
than -2, which would be systematically steeper than the observed distribution.
Also we here have abstained from modeling the transient 
aspect of real crypto-currencies. This allowed us to model a distribution
that does depend on the functional form of the abundance of new currencies.
An introduction that indeed should be expected to give an abundance of yet 
small currencies, and reduce the relative abundance of bigger currencies.
Finally, the social network aspect of emergent crypto-currencies may 
contribute to the exponent, as information and trust about small 
currencies may be localized in certain regions of the human social network 
\cite{Rosvall09}, which is potentially quantifiable by following transactions
of crypto-currencies, as well as information flows about the currencies 
\cite{Kristoufek2013,Kondor2014}.
 
Overall, our consideration serves to emphasize the crypto-currency
as a good model-system for the study of human folly, including the history-dependent 
randomness in assigning what is valuable and what has no value.
A consideration that should be at the heart of multiple aspects of 
social activity, social hierarchies, and thereby also be part of 
maintaining overall social order.

\end{document}